# Subwavelength focusing of all-dielectric surface waves

*Myun-Sik Kim,[†] Babak Vosoughi Lahijani,[†] Nicolas Descharmes,[†] Jakob Straubel,[‡] Fernando Negredo,[‡] Carsten Rockstuhl,[‡,§] Markus Häyrinen,[‖] Markku Kuittinen,[‖] Matthieu Roussey,[‖] and Hans Peter Herzig[†]*

[†]Optics & Photonics Technology Laboratory, Ecole Polytechnique Fédérale de Lausanne (EPFL), Neuchâtel, Switzerland

[‡]Institute of Theoretical Solid State Physics, Karlsruhe Institute of Technology, Karlsruhe 76131, Germany

[§]Institute of Nanotechnology, Karlsruhe Institute of Technology, Karlsruhe 76021, Germany

[‖]Institute of Photonics, University of Eastern Finland, P.O. Box 111, 80101 Joensuu, Finland.

**ABSTRACT:** Micro-sized spheres can focus light into subwavelength spatial domains: a phenomena called *photonic nanojet*. Even though well studied in three-dimensional (3D) configurations, only a few attempts have been reported to observe similar phenomena in two-dimensional (2D) systems. This, however, is important to take advantage of photonic nanojets in integrated optical systems. Usually, surface plasmon polaritons are suggested for this purpose, but they suffer notoriously from the rather low propagation lengths due to intrinsic absorption. Here, we solve this problem and explore, theoretically, numerically, and experimentally, the use of Bloch surface waves sustained by a suitably structured all-dielectric media to enable subwavelength focusing in an integrated planar optical system. Since only a low index contrast can be achieved while relying on Bloch surface waves, we perceive a new functional element that allows a tight focusing and the observation of a photonic nanojet on top of the surface. We experimentally demonstrate a spot size of $0.66\lambda$ in the effective medium. Our approach paves the way to 2D all-dielectric photonic chips for nano-particle manipulation in fluidic devices and sensing applications.

2D surface wave devices are a key asset in the miniaturization of compact optical systems, such as photonic integrated circuits and lab-on-a-chip devices. A frequently studied surface wave is the surface plasmon polariton (SPP) that propagates along the interface between a dielectric and a metal.[1] Since the field of SPPs is partially localized in the metal, SPPs are limited in their propagation length by the dissipation. This constitutes a severe limitation for many existing and potential applications, in particular, in integrated photonic devices. In that point of view, all-dielectric surface waves are much more appealing because they do not suffer from dissipation.

Two dielectric surface waves are known by now. One is called the Dyakonov wave, in reference to M. I. Dyakonov who made the first prediction.[2] It is a surface wave that propagates at the interface between two dielectric media where at least one of the two media should be anisotropic.[3] This requirement limits the fabrication possibilities and also the application perspective. The second is called a Bloch surface wave (BSW).[4] BSWs are supported at the edge of a truncated one dimensional (1D) periodic dielectric media, *i.e.*, a dielectric multilayer serving as 1D photonic crystals.[5] BSW are sustained in the frequency region of the localized photonic band-gap. The band-gap denies propagation into the half space containing the multilayer structure. Since the propagation constant of the mode is outside the light cone of the medium in the other half space, a surface mode is supported. At the edge of the truncated 1D photonic crystal, the thickness of the terminating layer is usually suitably tuned to let the BSW propagate at a frequency central to the band gap of the 1D photonic crystal. This improves its confinement. This last layer is called the defect layer. In an experiment however, the finiteness of the multilayer structure causes some radiation losses. That finite radiation loss is also important to excite BSWs by means of frustrated total internal reflection. Since BSWs can reside between isotropic dielectric media, their choice of materials and fabrication are much more flexible and a wide range of ambient materials are allowed. In addition, the use of lossless dielectric media guarantees a long propagation length. For instance, a BSW with a propagation length of 3.24 mm at λ = 1.558 μm has been reported.[6] Such promising characteristics of BSWs have inspired various studies in recent years. Devices with basic functionalities have been explored, *e.g.*, a prism demonstrating refraction of the propagating surface wave,[7] a grating for the generation of 2D Talbot images on the multilayer surface,[7] a plano-convex lens,[8] and waveguide components.[9-11] All these elements came in reach by adding a structured element layer on top of the stratified media. This locally changes the dispersion relation of the BSW and allows controlling the way it propagates. More advanced functionalities have also been investigated, *e.g.*, linear[12] and circular[13] grating couplers that couple the surface waves without a prism coupling setup like the Kretschmann configuration.[14] Furthermore, the research has been extended to demonstrate applications for enhanced fluorescence detection,[13] resonator,[15] and bio-sensing.[16] However, one of the most essential functionality of an optical system, a tight focusing, turned out to be non-trivial and has not yet been reported. It can be explained by the rather limited effective refractive index contrast ($\Delta n$) of the BSW sustained by the multilayer platforms when considered with and without the element layer. For fixed designs of focusing elements, the optical power is mainly depending on $\Delta n$ according to the lens maker's formula. In conventional optical systems, *e.g.*, those based on air and glass systems, the index contrast is approximately 0.5 whereas for most dielectric surface waves the index contrast is only in the order of 0.1. This causes a weaker optical power when conventional component designs are applied to the BSWs.[8] In this paper, we mitigate this problem and theoretically and experimentally investigate a component design that realizes a better spatial confinement of Bloch surface waves while fully respecting the intrinsic limitations imposed by the low index contrast.

A prime example of tight focusing using a simple component is the scattering of plane waves at dielectric microspheres. Mie scattering effect assists to generate a highly confined beam, named photonic nanojet,[17] which propagates over several wavelengths while maintaining a subwavelength beam size, *e.g.*, the full width at half maximum (FWHM) spot size smaller than one half of the wavelength for optimal conditions of the refractive index contrast and the size of the sphere with respect to the wavelength. Like this, such a simple element, *i.e.*, a microsphere, serves as the highest NA focusing lens in air. Most appealing of photonic nanojets is their support of various super-resolution applications, *e.g.*, nano-lithography,[18] single molecule sensing,[19] nano-particle detection,[20] and super-resolution imaging systems.[21-24] The majority of photonic nanojet studies are devoted to 3D systems. For 2D systems, surface plasmon polariton devices have been applied to attempt a tight focusing, *e.g.*, a gradient-index lens[25] and SPP photonic nanojets were generated on metal surfaces.[26,27] However, dielectric surface wave devices have not yet demonstrated to allow such a tight focusing. We approach this challenge here by developing new components enabling focusing 2D all-dielectric surface waves to sub-wavelength domains. Emphasis is put on studying the influence of the low index contrast on the photonic nanojet focusing and how to circumvent the weaker focusing using alternative geometry. We are driven by the idea to avoid complexity in the design and small feature sizes to continue to rely on standard fabrication processes.

In previous works,[6-13] multilayer platforms for the Bloch surface wave have been established from visible to near-infrared (IR) wavelengths. Here, a BSW platform operating at near-IR frequencies has been employed. The near-field distribution of a selected device has been measured with a custom scanning near-field optical microscope (SNOM). Effective modeling using a simplified 2D system and full-wave numerical 3D simulations of the actual structure provide further insights. We stress the fact that even though demonstrated here at a specific operational frequency, the scalability of Maxwell's equations allows to easily extrapolate all observed effects to different spectral regions, as long as materials with suitable dielectric properties are available.

- **METHODS AND EXPERIMENTAL DETAILS**

Figure 1 shows the schematic of our experimental setup. It includes the multilayer structures, a Kretschmann coupling configuration, and the SNOM probe. For the illumination we use a free space wavelength of $\lambda = 1.555$ μm.

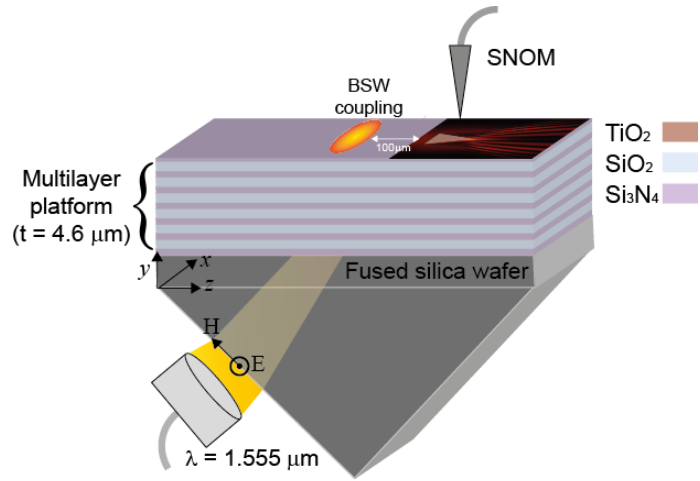

**Figure 1.** Schematic drawing of the experimental setup including the multilayer structure, Kretschmann coupling configuration, and the SNOM probe for near-field measurements. The multilayer platform consists of six periods of a layer made from $Si_3N_4$ and from $SiO_2$ with a thickness of 283 nm and 472 nm, respectively. A $Si_3N_4$ defect layer was deposited on top with a thickness of 50 nm. The element layer is 47 nm thick and made from $TiO_2$. This layer is patterned to form the focusing element.

The polarization is transverse electric (TE). The multilayer design has been conducted by calculating the band-gap diagram and dispersion curves using a transfer matrix method[5] and a numerical eigenmode solver CAMFR.[28] The designed multilayer stack consist of six periods of silicon nitride (Si$_3$N$_4$, $n_{Si3N4}$ = 1.94, $t$ = 283 nm) and silicon dioxide (SiO$_2$, $n_{SiO2}$ = 1.47, $t$ = 472 nm), where $t$ being thickness. An additional 50-nm-thick layer of Si$_3$N$_4$ is deposited on the top of the periodic layers as defect layer. The element layer is a 47-nm-thick titanium dioxide layer (TiO$_2$, $n_{TiO2}$ = 2.23), which is patterned to form the focusing element. The total thickness of the multilayer stack is approximately $t$ = 4.6 μm.

The dispersion curves with corresponding band-gap diagram of the current chip design is shown in Fig. 2(a). The electric field distribution throughout the multilayers at the operational wavelength of 1.555 μm is shown in Fig. 2(b), which verifies the efficient confinement of the BSW. The mode can be, however, excited by means of frustrated total internal reflection through the glass prism as performed in the Kretschmann configuration. Increasing the thickness of the defect layer causes the dispersion line to move towards the band edge near the glass light line. When adding the 47-nm-thick element layer made form TiO$_2$ on the top of the multilayer platform (1D photonic crystal + defect layer), the dispersion curve of the entire system shifts further to the glass light line.

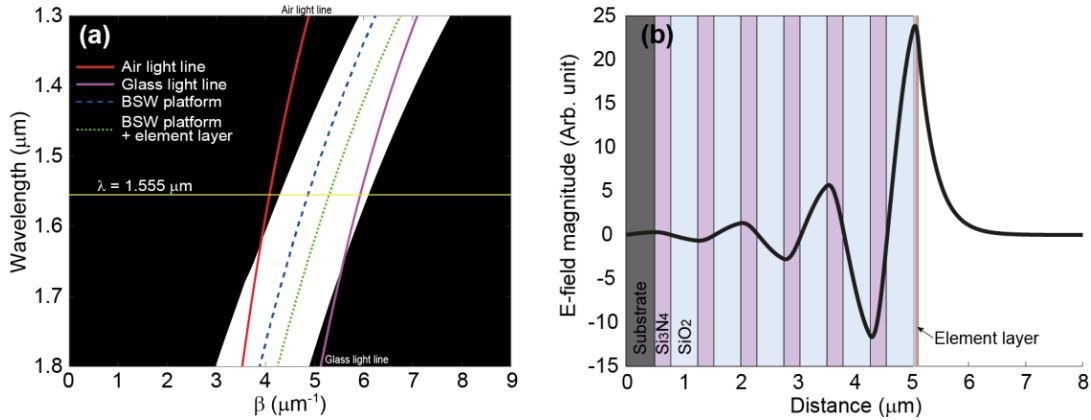

**Figure 2.** (a) Dispersion curves of the designed multilayer platform with/without the element layer and the band-gap diagram (the clear area is the localized photonic band-gap). (b) The electric field profile of the BSW throughout the multilayer structures at the operational wavelength of 1.555 μm.

For the operational wavelength, the dispersion curves of the platforms with/without the element layer causes an effective refractive index difference, as shown in Fig. 2(a) by blue dashed and green dotted lines within the band-bap. The effective refractive index of the BSW is defined by $n_{BSW}=c·\beta/\omega$, where $c$ being speed of light in vacuum, $\omega$ being the angular frequency, and $\beta = k·\sin\theta$ being the wave number of the surface mode with $k$ the wave number in the glass substrate and $\theta$ the incident angle in the Kretschmann configuration.[8] From Fig. 2(a), the effective refractive indices of the platform with and without the element layer can be extracted as $n_{eff1} = 1.3$ and $n_{eff2} = 1.2$, respectively, resulting in an index contrast $\Delta n_{eff} = 0.1$. Accordingly, the effective wavelength of the BSW ($\lambda_{BSW}$) propagating on the multilayer platform without the element layer is the free space wavelength times the reciprocal of $n_{eff2}$. The band-gap and the substrate dispersion line limit the maximum index contrast. In theory and while ignoring other restrictions of fabrications and coupling conditions, high index substrate materials, *e.g.*, amorphous silicon with an $n = 3.48$ can expand the $\Delta n_{eff}$ up to approximately 0.38 with a similar multilayer stack design. However, it is impossible to achieve the optimal index contrast for the conventional photonic nanojet generation. The location and intensity of photonic nanojets strongly depend on the refractive index contrast between the sphere and its surrounding medium, as well as its size. Since there is no basic relation, accurate location and spot size of photonic nanojets are usually obtained by rigorous numerical methods,[17,29] where the optimal index contrast to acquire a subwavelength spot size is found to be in the range between 0.5 to 1.

Since the index contrast of the BSW systems does not meet the optimal condition for the conventional photonic nanojet based on spherical geometries, alternative designs have to be found to squeeze the light into subwavelength spatial domains. It is known that non-circular shapes, for instance pyramid shapes[30] or square pillars,[31] can generate photonic nanojet spots as well.

Therefore, we explored such elements. We consider standard fabrication processes by avoiding small feature sizes and fixed the critical dimensions of the considered structure to 10 μm. Structures with these dimensions can be easily patterned by conventional lithographic methods. Avoiding complex designs, we investigate simple geometries, such as, circle, ellipse and triangle. The performance of each geometry has been assessed by studying at first the light propagation numerically. Preliminary simulations based on fast and resource efficient 2D simulations. For the final design, selected full 3D simulations considering the exact geometry were done. All simulations have been performed using a commercial time domain solver that is based on the finite integration technique as implemented in CST Microwave Studio. In the preliminary 2D simulations, we consider a scalar situation and light propagation in a medium characterized by only the effective refractive indices of the focusing element and the effective surrounding medium. We do not include the multilayer stacks as is done in full 3D modeling. This 2D model mimics the propagation of the BSW along the surface.

- **RESULTS AND DISCUSSION**

We first verify how much we can reduce the focal spot by elongating the conventional circular design to a prolate ellipse (see Supporting Information). The axis length of the ellipse in one dimension was considered as the critical dimension of the structure and was fixed to 10 μm. The second axis length of the ellipse was changed from 10 μm to 25 μm with an interval of 5 μm. However, the low index contrast ($\Delta n_{\mathrm{eff}} = 0.1$) forces the circular and prolate ellipse designs to act rather as a double convex lens and do not yield a photonic nanojet. Small refraction angle from the illumination-side surface lowers the NA of the system and a tight focusing was not observed. Therefore, the confinement of light in the focal spot was weaker than that of conventional photonic nanojets. For instance, the smallest focal spot with an elliptical design, shown in the Supporting

Information in Fig. S1(d), has a FWHM size of 1.25 μm, which is approximately 0.96 $\lambda_{BSW}$. This is far away from the typical spot size of conventional photonic nanojet, *i.e.*, slightly larger than half the wavelength.

To further enhance the light confinement towards a subwavelength size, we introduce a non-circular design and consider an isosceles triangle. With such a geometry, we return eventually to the prototypical structure that allows inducing diffraction free beams in a 2D geometry. The isosceles triangle introduces two spatial frequencies that are linked to the index contrast and the geometrical dimensions. These two plane waves share the same propagation constant and have oppositely signed wave vector components in the transverse direction. They interfere and eventually cause a standing wave pattern that is invariant along the propagation direction. The finiteness of the isosceles triangle limits, obviously, the spatial extent where the beam is diffraction free, comparable to the design constraints of an axicon. By changing the height of the isosceles triangle we can change the spatial extent in which the beam is free of diffraction, limiting it closer to the top of the triangle. The two counter-propagating waves, in the limiting case, can be focused to exactly half a wavelength, which promises to obtain a spatial localization of the field close to the tip of the triangle approaching this fundamental lower bound. This design approach has been first numerically investigated using 2D simulations. In our analysis we fixed the base to be $w = 10$ μm and varied the height $h$ of the isosceles triangle from 5 μm to 20 μm with an interval of 5 μm.

Figure 3 shows the intensity distributions of 2D simulation results for different aspect ratios ($h/w$). Note that the intensity is the square of the electric field amplitude and the incidence is the *x*-polarized plane wave in these 2D simulations. When the $h/w$ ratio grows, the focal spot gradually becomes smaller until it is buried in the element [see Fig. 3(d)]. The best confinement is found for $h/w = 1.5$, as shown in Fig. 3(c). Similar to the photonic nanojet spots, the focal spot arises just at the top of the isosceles triangle. The FWHM spot size reaches down to 950 nm, *i.e.*, 0.73$\lambda_{BSW}$. The behavior is in agreement with the physical understanding developed above. For a shallow height, the induced transverse momentum given to the two plane waves is quite small. This causes the modulation of the intensity in the transverse direction to be spatially extended, *i.e.*, there is no tight focusing. Increasing the height will increase the induced transverse momentum. This squeezes the spatial extent. But the structure is neither infinitely extended nor periodically arranged. This limits the extension in the propagation direction in which the beam can be considered as free of diffraction, *i.e.*, it restricts the longitudinal size of the photonic nanojet. Numerically we observe that an optimum is met at the ratio of $h/w$ = 1.5. There, a tight focusing above the tip of the triangle is obtained. This optimum design eventually is the structure we are looking at in the experiments described further below.

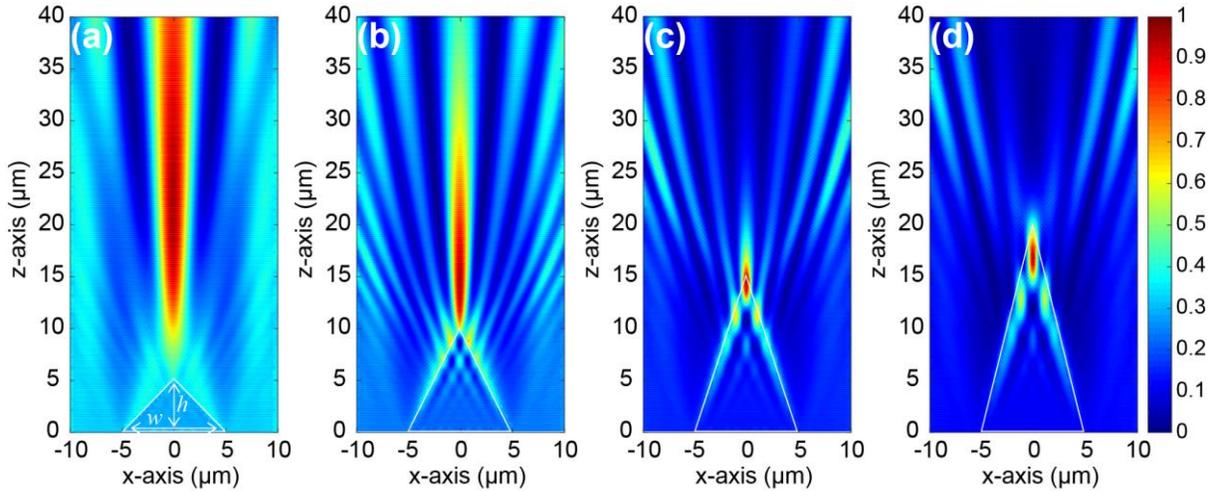

**Figure 3.** 2D simulation results of the intensity distributions, at $\lambda$ = 1.555 μm, for photonic nanojet devices for BSWs based on isosceles triangles. While the base of $w$ = 10 μm has been fixed and the height $h$ varied from 5 μm to 20 μm: (a) - (d) $h$ = 5, 10, 15, 20 μm, respectively. (c) For the aspect ratio $h/w$ equal to 1.5, the best confinement is obtained, where the FWHM spot size equals

$0.73\lambda_{BSW}$. (d) When the aspect ratio further increases, the focal spot is buried in the element. The intensity is normalized to the maximum of each figure.

For this final design, studied in Fig. 3(c), we also apply a full 3D simulations as described above to verify the validity of 2D results (for details see Supporting Information). The results of the 3D simulations are shown in Fig. 4 along with the experimental results for comparison.

For the experimental verification, we have fabricated the funnel device whose fields has been studied in Fig. 3(c), *i.e.*, the isosceles triangle of $w$ = 10 μm and $h$ = 15 μm in the element layer. Multilayers of $Si_3N_4$ and $SiO_2$ are deposited by using plasma-enhanced chemical vapor deposition (PECVD) on a fused silica wafer. The element layer ($TiO_2$) layer was deposited by atomic layer deposition (ALD). The patterning of $TiO_2$ layer was performed by electron-beam lithography (EBL) and subsequent inductively coupled plasma reactive ion etching processes (ICP-RIE). The details of the fabrication can be found in Supporting Information.

Using the Kretschmann coupling configuration shown in Fig. 1, the incident laser beam with TE polarization and a free space wavelength of $\lambda$ = 1.555 μm is successfully coupled to the top of the multilayer 100 μm away from the fabricated isosceles triangle element. The illumination is a weakly focused Gaussian beam with a FWHM size of 25 μm. Since it is coupled at the focal plane of the illumination lens, the propagating surface wave exhibits a planar wavefront that is not diverging. This imitates a collimated beam bound to the surface, which is the experimental counterpart of the plane wave illumination considered in the simulation. The SNOM setup employs a commercial metal-coated fiber tip (200-nm aperture) to scan over the funnel device. A scan in the *xz*-plane over a spatial region of 10 μm by 20 μm with a scan step of 100 nm in both directions has been performed. Figure 4(b) shows the measured near-field intensity distribution. It shows an excellent agreement with the numerical results shown in Figs. 3(c) and 4(a) corresponding to the 2D and 3D simulations, respectively. For further analysis, we extract and plot the intensity profiles at $z = 15$ μm along the *x*-axis from numerical and experimental results. The direct comparison is shown in Fig. 4(c), where all profiles are normalized to have the same peak value. Except the negligible deviations in the side lobes, the main lobe profile shows an excellent match. The measured FWHM spot size is 858 nm, *i.e.*, $0.66\lambda_{BSW}$. Fabrication and experimental errors, such as, an alignment error and imperfect illumination conditions, may cause the deviations in the side lobes.

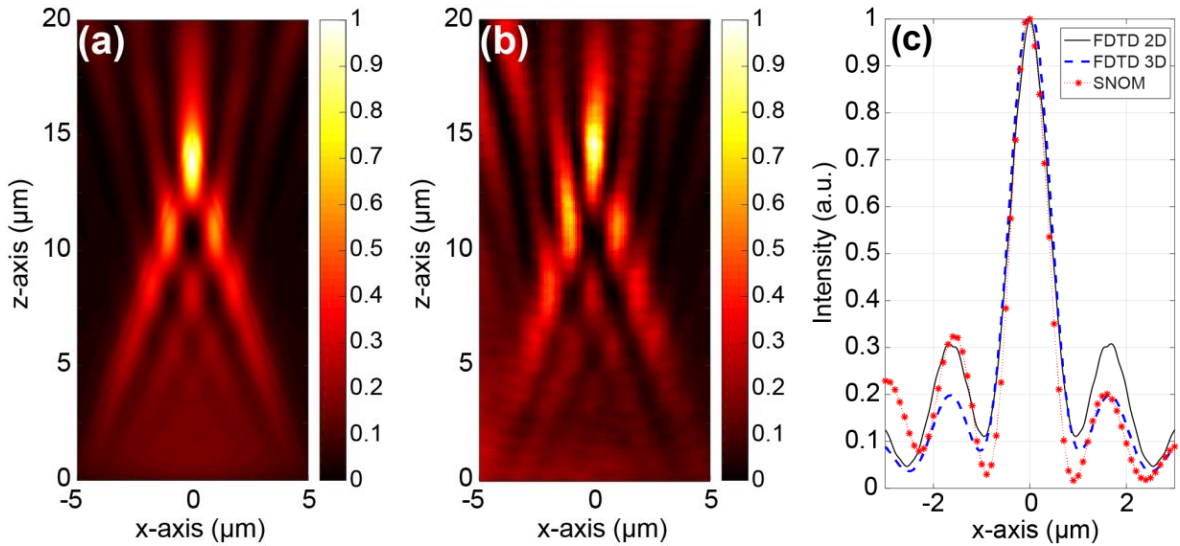

**Figure 4.** (a) The 3D simulation result for the *x-z* intensity (that is electric field amplitude squared) distribution. (b) The measured near-field intensity distribution of the fabricated funnel device. (c) The *x*-axis profiles at $z = 15$ μm of the 2D (dark solid line) and the 3D (blue dashed line) simulations, and the measured intensity (red dotted-asterisk line). Results are extracted from Figs. 3(c), 4(a), and 4(b), respectively.

▪ **CONCLUSIONS**

We have investigated both theoretically and numerically how to focus all-dielectric surface waves into a subwavelength spatial domain using near-IR Bloch surface wave of free space wavelength λ = 1.555 μm. We considered a simple component such as a microsphere that generates a subwavelength focus via the photonic nanojet effect. However, the available material platforms to host the BSW provide as an intrinsic limitation only a rather low index contrast ($\Delta n_{\text{eff}}$ = 0.1 for the current multilayer design) to steer the propagation of the BSW. This denies achieving the optimal index contrast condition for the conventional photonic nanojet that requires a $\Delta n$ to be in the range of 0.5 to 1. By means of simplified 2D simulations, it has been demonstrated that circular and elliptical designs do not provide spot sizes comparably small to those of conventional photonic nanojets. To enhance the spatial confinement, we suggested a novel device concept that relies on an isosceles triangle. It introduces two spatial frequencies that cause a pattern free of diffraction for a shallow height that is, however, not strongly localized in space. Increasing the height of the isosceles triangle enhances the localization but lowers simultaneously the diffraction free extent of the propagation direction. The optimal design has been obtained by an isosceles triangle of the width $w$ = 10 μm and the height $h$ = 15μm. For this final design, a full 3D simulation verifies the validity of 2D simplified simulation results. By using a SNOM measurement, we experimentally verify the performance of the fabricated device, demonstrating the FWHM spot size of 858 nm (= $0.66\lambda_{\text{BSW}}$). Considering fabrication and experimental deviations, the results show an excellent agreement with the 2D and 3D simulation results.

Thanks to their all-dielectric nature, BSW platforms pave the way for integrated photonic systems with very large propagation lengths. The demonstrated subwavelength focusing of the surface waves can fill the vacancy of the all-dielectric surface wave components for single molecule sensing,[19] nano-particle detection,[20] trapping and sensing,[32-34] and all-optical switching.[35]

- **ACKNOWLEDGMENT**

This research is supported by the Swiss National Science Foundation (SNSF FN200020-135455), Finnish Funding Agency for Innovation (Project Tekes) FiDiPro NP-NANO (40315/13), and the Deutsche Forschungsgemeinschaft (DFG) through CRC 1173. J.S. and F.N. also acknowledge support from the Karlsruhe School of Optics and Photonics (KSOP).